\documentclass[a4paper,11pt]{article}
\usepackage{pos}
\usepackage{slashed}

\title{Stability of homogeneous chiral phases against inhomogeneous perturbations in 2+1 dimensions}
\ShortTitle{Stability of homogeneous phases against inhomogeneous perturbations in 2+1 dimensions}

\author*[a]{Marc Winstel}
\author[a]{Laurin Pannullo}

\affiliation[a]{Institut für Theoretische Physik, Goethe Universit\"at Frankfurt am Main \\
	Max-von-Laue-Str. 1, 60438 Frankfurt am Main, Germany}

\emailAdd{winstel@itp.uni-frankfurt.de}
\emailAdd{pannullo@itp.uni-frankfurt.de}

\abstract{

In this work, inhomogeneous chiral phases are studied in a variety of Four-Fermion and Yukawa models in $2+1$ dimensions at zero and non-zero temperature and chemical potentials. Employing the mean-field approximation, we do not find indications for an inhomogeneous phase in any of the studied models. We show that the homogeneous phases are stable against inhomogeneous perturbations.  At zero temperature, full analytic results are presented.	

}

\FullConference{%
	The 39th International Symposium on Lattice Field Theory,\\
	8th-13th August, 2022,\\
	Rheinische Friedrich-Wilhelms-Universität Bonn, Bonn, Germany
}
\usepackage{hyperref}
\usepackage[capitalise]{cleveref}
\usepackage{graphicx}
\usepackage{subcaption}
\usepackage{mathtools}
\usepackage{amsmath}
\usepackage{xcolor}
\usepackage{dsfont}
\usepackage{bbold}
\usepackage[utf8]{inputenc}
\usepackage[ngerman, english]{babel}
\usepackage{xstring}
\usepackage{accents}
\usepackage{upgreek}

\renewcommand{\ref}[1]{(\ref{#1})}

\newcommand{\N}{\ensuremath{N}}
\newcommand{\ii}{\ensuremath{\mathrm{i}}}

\providecommand{\Rcite}[1]{%
	\begingroup
	\def\tempx{0}%
	\StrCount{#1}{,}[\tempx]%
	\ifnum\tempx > 0 
	Refs.~%
	\else
	Ref.~%
	\fi
	\endgroup
	\cite{#1}%
}

\DeclareMathOperator{\Q}{Q}
\definecolor{corr}{rgb}{0,0,1}

\begin{document}
	\maketitle
\section{Introduction}
Inhomogeneous chiral condensates, where both chiral symmetry and translational invariance are spontaneously broken by non-vanishing, spatially dependent $\langle \bar{\psi} \psi \rangle(\mathbf{x})$, have been observed in several model studies, most-prominently the Gross-Neveu (GN) model \cite{Gross:1974jv} and related models in 1+1 dimensions (see \Rcite{Thies:2003kk, Thies:2006ti, Schon:2000he, Thies:2020ofv, Ebert:2011rg, Thies:2022kuv, Khunjua:2017khh}), but also in $3+1$-dimensional, low-energy effective models for QCD (see \Rcite{Buballa:2014tba} for a review and, e.g., \Rcite{Lakaschus:2020caq, Buballa:2020xaa, Carignano:2019ivp, Buballa:2018hux, Nickel:2009wj, Dumm:2021vop, Andersen:2018osr}). 
Albeit these phases have been found in the mean-field approximation, i.e., neglecting bosonic quantum fluctuations, it is discussed in recent literature \cite{Lenz:2020bxk, Stoll:2021ori, Ciccone:2022zkg, Lenz:2021kzo, Lenz:2021vdz, Nonaka:2021pwm} whether such an inhomogeneous chiral phase might leave its imprint when studying the full quantum field theories.
In $2+1$ dimensions the situation seems to differ from even spacetime dimensions.
Even though inhomogeneous condensates can be favored at finite regulator values in the $2+1$-dimensional GN model (depending on the chosen regularization scheme), the inhomogeneous phase vanishes in the renormalized limit \cite{Winstel:2019zfn,Buballa:2020nsi, Narayanan:2020uqt, Winstel:2021yok,Pannullo:2021edr}. 
In this work, we analyze the stability of homogeneous condensates in a variety of Four-Fermion (FF) models and recover expressions similar to those derived in \Rcite{Buballa:2020nsi}. 
No indications for an inhomogeneous phase are observed in all of the discussed models. 
The homogeneous ground states are stable against inhomogeneous perturbations after renormalization.
We argue that this statement, observed in these bosonized FF models, also remains valid when studying corresponding Yukawa models\footnote{By corresponding Yukawa models we mean those Yukawa models that are built out of the bosonized FF action by adding kinetic terms and self-interactions for the auxiliary bosonic fields. Details are discussed in Sec.~\ref{sec:Yukawa}.}.
Numerical minimizations of the lattice actions allow us to determine the preferred ground states at finite lattice spacing. 
The results obtained by the two different methods are in agreement, as both approaches show no indications for the existence of an inhomogeneous phase in the continuum.
\section{Detecting inhomogeneous phases via the momentum dependence of the bosonic two-point function\label{sec:stability}}
In Sec.~\ref{sec:res}, we study a variety of FF models, which feature different interaction channels and chemical potentials. 
In order to introduce our method, the stability analysis, we define a general FF model, which can be reduced to the later mentioned actions by choosing certain values for the model parameters. 
The action of the FF model in Euclidean spacetime is given by
\begin{equation}
	S_{\mathrm{FF}}[\bar{\psi},\psi] = \int \mathrm{d}^3x  \ \left\{ \bar{\psi}(x)\left(\slashed{\partial}+ \gamma_0 \mu + \gamma_0 \gamma_{45} \mu_{45}\right)  \psi(x) -  \left[ \sum_{j=1}^{16} \tfrac{\lambda_j}{2 \N } \left(\bar{\psi}(x)\,  c_j\, \psi(x)\right)^2  \right]\right\}, \label{eq:FFmodel} 
\end{equation} 
where $\psi$ contains $2\N$ four-component spinors ($\N$ identical spinors with isospin up/down respectively) without a bare mass term for the spinors. 
The Dirac matrices are chosen as reducible $4\times4$ representations of the $2+1$-dimensional Euclidean Clifford algebra (for details see, e.g.~\Rcite{Buballa:2020nsi, Gies:2009da, Pisarski:1984dj, Pannullo:2021edr}). 
The matrices $c_j$ are elements of  
\begin{equation}
C = \{c_j\}_{j=1,\ldots, 16} = \{\mathds{1}, \ii\gamma_4, \ii\gamma_5, \gamma_{45}, \vec{\tau}, \ii\vec{\tau}\gamma_4, \ii\vec{\tau}\gamma_5, \vec{\tau}\gamma_{45}\}, \label{eq:dirac_basis}
\end{equation}
where $\vec{\tau}$ is the vector of Pauli-matrices acting on the isospin degrees of freedom. 
The integration over the temporal direction $x_0$ is restricted to the compact interval $[0,\beta)$ at temperature $T=\tfrac{1}{\beta}$. 
The coupling of the FF terms is denoted by $\lambda_i$. The baryon chemical potential is denoted as $\mu$, while $\mu_{45}$ describes the chiral chemical potential\footnote{In general, our formalism allows us to include several chemical potentials. 
For simplicity, we will only discuss $\mu$ and $\mu_{45}$ throughout this work.}. 
After bosonization with one auxiliary bosonic field $\phi_j$ for each channel and integration over the fermion fields, one obtains an equivalent effective action given by
\begin{align}
\frac{S_{\mathrm{eff}}[\vec{\phi}]}{\N} &= \int \mathrm{d}^3x \ \  \sum_i \frac{\phi_i^2(\mathbf{x})}{2 \lambda_i}  \ -  \  \mathrm{Tr} \ln  \Q, \label{eq:Seff}
 \\
\Q &= \slashed{\partial}+ \gamma_0 \mu + \gamma_0 \gamma_{45} \mu_{45} + \sum_{j} \, c_j \, \phi_j(\mathbf{x}),
\end{align}
where $\mathrm{Tr}$ is a functional trace over the spacetime coordinates as well as over the internal fermionic degrees of freedom. 
The 16 real bosonic fields $\vec{\phi}$ are restricted to depend on the two spatial coordinates $\mathbf{x} = (x_1, x_2)$. They are related to fermion bilinears via the identities 
\begin{equation}
	\langle \phi_j\rangle = - \tfrac{\lambda_j}{\N} \langle \bar{\psi} c_j \psi\rangle, \quad j = 1, \ldots, 16. \label{eq:Ward}
\end{equation}
Suppressing the bosonic fluctuations in the path integral, i.e., taking $\N \rightarrow \infty$, the computation of observables is equivalent to their evaluation on the global minima of the effective action with respect to the bosonic fields $\vec{\phi}(\mathbf{x})$. 
Finding the global minima $\vec{\phi}(\mathbf{x}) = \vec{\Phi}(\mathbf{x})$ of the effective action \eqref{eq:Seff} for all possible field configurations is an extremely challenging task, both analytically and numerically. 
Thus, we simplify the search for inhomogeneous condensates by analyzing the stability of the homogeneous ground state $\vec{\phi}(\mathbf{x}) = \vec{\bar\Phi}$ against inhomogeneous perturbations (see \Rcite{Koenigstein:2021llr} for a detailed discussion and test of the method). 
Such an analysis was already applied in \cite{Buballa:2018hux, Carignano:2019ivp, Nakano:2004cd, Buballa:2020nsi}. 
The analysis requires a homogeneous expansion point, i.e.,
\begin{equation}
\vec{\phi}(\mathbf{x}) = \vec{\bar{\phi}}  + \delta \vec{\phi} (\mathbf{x}),
\end{equation}
where $\delta \vec{ \phi}(\mathbf{x})$ describes the spatially dependent inhomogeneous perturbation around the homogeneous expansion point $\vec{\bar{\phi}}$. 
Then, the effective action is expanded in terms of the perturbation $\delta \vec{ \phi}(\mathbf{x})$. 
Here, the zeroth order correction corresponds to $S_{\mathrm{eff}}[\vec{\bar{\phi}}]$. 
The first order correction $S_\mathrm{eff}^{(1)}$ vanishes due to the gap equation when evaluated at the homogeneous minima, e.g.~the homogeneous ground state $\vec{\bar{\phi}} = \vec{\bar{\Phi}}$. 
Thus, we will not explicitly present $S_\mathrm{eff}^{(1)}$ in the following. The second order correction evaluated in momentum space is given by
\begin{align}
\frac{S_\mathrm{eff}^{(2)}[\vec{\bar{\phi}}]}{N} &=  \frac{\beta}{2}  \int \frac{\mathrm{d}^2 q}{(2\uppi)^2}\, \left[\sum_{i, j}\, \delta \tilde{\phi}_i^*(\mathbf{q})\, \delta\tilde{\phi}_j(\mathbf{q})\, \left(\delta_{i,j}\,\lambda^{-1}_i+\Gamma_{\phi_i \phi_j}^F (\mathbf{q}^2)\right) \right] , \label{eq:Seff2}  \\
&\Gamma^F_{\phi_i \phi_j} =  \int \frac{\mathrm{d}^3p}{(2\uppi)^3}\,  \mathrm{tr}\left(c_i\, \bar{Q}^{-1}(p+q)\, c_j\, \bar{Q}^{-1}(p)\right), \label{eq:ferm_curvature}
\end{align}
where $\mathbf{q}$ is the momentum of the inhomogeneous perturbation and  $q = \left(0, \mathbf{q}\right)$. 
Eq.~\eqref{eq:ferm_curvature} is the contribution of the fermion $1$-loop diagram with two amputated bosonic legs and, correspondingly, contains two fermion propagators with the homogeneous fields $\vec{\bar{\phi}}$, i.e.~the inverse of  
\begin{equation}
	\bar{Q}(p) = \slashed{p} + \gamma_0 \mu + \gamma_0 \gamma_{45} \mu_{45} + \sum_{j} \, c_j \, \bar{\phi}_j.
\end{equation}
To compute the curvature in each field variable, one has to diagonalize the quantity in parentheses in Eq.\ \eqref{eq:Seff2} with respect to $\delta \vec{\tilde{\phi}}(\mathbf{q})$ by finding a suitable basis of bosonic fields $\delta \vec{ \tilde{\varphi}}(\mathbf{q})$. 
If it is possible to transform to such variables, one obtains a two-point function $\Gamma_{\varphi_j}^{(2)}$ for each corresponding field $\varphi_j,\, j=1, \ldots, 16$. 
The detection of an inhomogeneous condensate, which yields a lower effective action than $S_\mathrm{eff}[\vec{\bar{\Phi}}]$ for a given $\mu$ and $T$, corresponds to obtaining a two-point function $\Gamma_{\varphi_j}(\mathbf{q}) < 0$ with $\mathbf{q} \neq 0$ when the two-point function is evaluated at $\vec{\phi} = \vec{\bar{\Phi}}$.
\subsection{Generalization to Yukawa models\label{sec:Yukawa}}
In general, our conclusions drawn from the bosonic two-point functions of the bosonized FF models can be generalized to corresponding Yukawa models, where the auxiliary bosonic fields are promoted to dynamical fields $\chi$ with $2n$-point self-interactions and kinetic terms, i.e.,
\begin{equation}
S_{\mathrm{eff}, Y} [\vec{\chi}] =  \frac{S_{\mathrm{eff}}[h\vec{\chi}]}{\N} + \int \mathrm{d}^3 x \left[ \frac{1}{2 \N} \left(\partial_\nu \vec{\chi}(\mathbf{x})\right) \left(\partial_\nu \vec{\chi}(\mathbf{x})\right) + \sum_{n}\frac{\kappa_n}{\N} \left( \sum_{j} \chi_j^2(\mathbf{x})\right)^{n}\right],  \label{eq:Yukawa}
\end{equation}
with the Yukawa-coupling $h$ and the couplings $\kappa_n$ of the $2n$-self interactions . 
The kinetic term yields a monotonically increasing contribution $\sim \mathbf{q}^2$ to the integrand in Eq.\ \eqref{eq:Seff2}, while the bosonic self-interaction gives a $\mathbf{q}$-independent offset.
These additional terms cannot cause a FF model without an instability to develop an instability.
Details will be discussed in an upcoming publication. 
\section{Results\label{sec:res}}
Before repeating the analysis for other FF models we shortly summarize the GN model results obtained in \Rcite{Buballa:2020nsi}. 
The two-point function of the GN model or analogous expressions are then identified for a variety of FF models. 
The FF models can all be derived from \eqref{eq:FFmodel} by setting certain couplings $\lambda_i$ to zero, while the non-zero couplings are set to $\lambda$. 
The result regarding the stability of the homogeneous condensates against inhomogeneous perturbations implicitly also holds for the corresponding Yukawa models as discussed in Sec.\ \ref{sec:Yukawa}. 
We set the chiral chemical potential $\mu_{45}=0$ if not explicitly stated otherwise.
\subsection{GN model}
The effective action of the GN model \cite{Gross:1974jv} is given by 
\begin{align}
\label{eq:bos_action} \frac{S_{\text{eff}}[\sigma]}{\N} = \frac{1}{2 \lambda} \int \mathrm{d}^3x \, \sigma^2(\mathbf{x}) - \mathrm{Tr}\ln(\slashed{\partial} + \gamma_0 \mu + \sigma(\mathbf{x})),
\end{align}
where $\sigma$ is the auxiliary bosonic field, whose expectation value is proportional to the fermion bilinear $\langle\bar{\psi} \psi\rangle$ via \cref{eq:Ward}. 
The expansion around the homogeneous ground state $\bar{\sigma} = \bar{\Sigma}$, as discussed for arbitrary $\bar{\sigma}$ in \Rcite{Buballa:2020nsi}, results in a vanishing first order correction $S_{\text{eff}}^{(1)}$. 
The second order correction evaluated in momentum space is given by
\begin{equation}
\frac{S_{\text{eff}}^{(2)}}{\N}=\frac{\beta}{4} \int \frac{\mathrm{d}^2q}{(2\uppi)^2} |\delta\tilde\sigma(\mathbf{q})|^2 \, \Gamma_\sigma^{(2)}(\mathbf{q}^2) 
\end{equation}
with 
\begin{equation}
\Gamma_\sigma^{(2)}(\mathbf{q}^2) =
\frac{1}{2\lambda}-\ell_1 +L_2(\mathbf{q}^2, \bar{\sigma}, \mu)   \label{eq:prot_2P}
\end{equation}
and 
\begin{equation}
	L_2(\mathbf{q}^2, \bar{\sigma}, \mu) =-\frac{1}{2}(\mathbf{q}^2+4\bar\sigma^2)\ell_2(\mathbf{q}^2, \bar{\sigma}, \mu),
\end{equation}
where $\ell_1$ is a linearly diverging integral (see \Rcite{Buballa:2020nsi} for the definitions of $\ell_1$ and $\ell_2$) and has to be regulated, e.g., with the Pauli-Villars regularization. The integral $\ell_2$ is finite, but, in general, is often also regulated for consistency\footnote{This is an arbitrary choice, which is however reasonable as it leaves mathematical relations between the unregulated integrals intact. Such a prescription is, e.g., applied in 3+1 dimensions, where FF models are non-renormalizable. It is also suitable to compare with lattice regularizations where anyhow all quantities are regulated. }. 
The coupling constant $\lambda$ can be tuned using the condensate in the vacuum such that $\tfrac{1}{2\lambda} - \ell_1 < \infty$ for all $\mu$ and $T$. 
A main finding of \Rcite{Buballa:2020nsi} is that $\Gamma_\sigma^{(2)}$ is a monotonically increasing function of $q = |\mathbf{q}|$ for all values of $\mu$ and $T$ after renormalization. 
For $T=0$, one can derive the monotonic behavior analytically, i.e., one finds 
\begin{equation}
L_2\big|_{T=0} = \begin{cases}
\frac{(q^2+4\bar\sigma^2)}{2\uppi q}\arctan\left(\frac{q}{2|\bar{\sigma}|}\right) & \text{if } \mu^2 < \bar{\sigma}^2 \\
\frac{(q^2+4\bar\sigma^2)}{2\uppi q}\arctan\left(\frac{\sqrt{q^2/4 - \mu^2 + \bar{\sigma}^2}}{|\mu|}\right) & \text{if } \mu^2 > \bar{\sigma}^2 \\
0 & \text{if } \mu^2 > \bar{\sigma}^2 + \frac{q^2}{4} \\
\end{cases}.
\end{equation} 
As a result, $\Gamma_\sigma^{(2)}(\mathbf{q}^2) \geq 0$ and no indications for an inhomogeneous phase can be found. 
One is left with the phase diagram derived in \Rcite{Klimenko:1987gi}. 
As discussed in \Rcite{Buballa:2020nsi}, one obtains $\Gamma_\sigma^{(2)}(\mathbf{q}^2) < 0$ at finite $\mathbf{q} \neq 0$ when studying finite regulator values, e.g., lattice spacings $a$, for certain regularization schemes. 
We refer to \Rcite{Buballa:2020nsi} for numerical results in the GN model.
In the following, we will identify analytical structures similar to $L_2(\mathbf{q}^2, \bar{\sigma}, \mu)$ in the other FF models. 
\subsection{$U(4\N)$ invariant model \label{sec:U4}}
A model, which -- in contrast to the GN model -- features a continuous chiral $U(4\N)$ symmetry\footnote{One free massless fermion field is invariant under $U(2)$ chiral symmetry transformations. Studying $N$ fermion fields with an isospin degree of freedom enlarge the symmetry group to $U(4\N)$ accounting to rotations in these spaces.} has the effective action
\begin{align}
\frac{S_{\text{eff}}[\sigma, \eta_4, \eta_5, \eta_{45}]}{\N}= \int \mathrm{d}^{3}x\, \tfrac{\rho^2}{2 \lambda} - \mathrm{Tr} \ln\left(\slashed{\partial}  + \gamma_0\mu + \sigma + \ii \gamma_4 \eta_4 + \ii \gamma_{5} \eta_5 + \gamma_{45} \eta_{45}\right),
\end{align}
where $\sigma, \eta_4, \eta_5, \eta_{45}$ are the auxiliary bosonic fields corresponding to fermionic bilinears, as discussed in Eq.~\eqref{eq:Ward}, and $\rho^2 = \sigma^2 + \eta_4^2 + \eta_5^2+ \eta_{45}^2$. 
A non-vanishing expectation value of $\eta_{45} $ indicates spontaneous parity breaking, in contrast to the other three fields, where a non-vanishing expectation value indicates spontaneous chiral symmetry breaking. 
Considering homogeneous fields $\left(\sigma, \eta_4, \eta_5, \eta_{45}\right) = \left(\bar{\sigma}, \bar{\eta}_4, \bar{\eta}_5, \bar{\eta}_{45}\right)$ one can rotate $\bar{\eta}_4 = \bar{\eta}_5 = 0$ through the $O(3)$ rotational symmetry of the model\footnote{This symmetry is linked to the $U(2)$ chiral symmetry. The field $\eta_{45}$ is not connected to the other fields via such a symmetry transformation, as the bilinear $\bar{\psi} \gamma_{45} \psi$ remains invariant under $U(2)$.}. One finds additionally $\bar{\eta}_{45} = 0$ for all $\mu$ and $T$ by explicitly minimizing $S_{\text{eff}}[\bar{\sigma}, 0, 0, \bar{\eta}_{45}]$.
With $\left(\sigma, \eta_4, \eta_5, \eta_{45}\right) = \left(\bar{\Sigma}, 0, 0, 0\right)$ as the expansion point one obtains a vanishing first order correction  $S_{\mathrm{eff}}^{(1)}$. The second order correction is then given by
\begin{align}
&\frac{S_{\text{eff}}^{(2)}}{\N} = \frac{\beta}{4} \int \frac{\mathrm{d}^2q}{(2\uppi)^2} \sum_{\phi\in \{\sigma,  \eta_4, \eta_5, \eta_{45}\}} |\delta\tilde\phi(\mathbf{q})|^2\, \Gamma_\phi^{(2)}(\mathbf{q}^2),\\
\Gamma^{(2)}_{\eta_{45}}= \Gamma^{(2)}_\sigma &=  \frac{1}{2\lambda}-\ell_1+ L_2(\mathbf{q}^2, \bar{\sigma}, \mu),\quad \label{eq:2point_76}
\Gamma^{(2)}_{\eta_4} = \Gamma^{(2)}_{\eta_5} =  \frac{1}{2\lambda}-\ell_1-\frac{1}{2}\mathbf{q}^2\ell_2(\mathbf{q}^2, \bar{\sigma}, \mu^2),
\end{align}
where the two-point functions $\Gamma^{(2)}_{\eta_{45}}, \, \Gamma^{(2)}_\sigma$ are equal to Eq.\ \eqref{eq:prot_2P}, while $\Gamma^{(2)}_{\eta_4}, \, \Gamma^{(2)}_{\eta_5}$ differ slightly in the prefactor of $\ell_2$ but are also monotonically increasing functions of $q$.
However, studying the model at finite lattice spacing $a$, as relevant for $3+1$-dimensional, non-renormalizable FF models (compare \cite{Nickel:2009wj, Carignano:2019ivp, Pannullo:Lattice2022}), one obtains negative values when evaluating the two-point functions \eqref{eq:2point_76} at finite $\mathbf{q} \neq 0$ in a certain ($a$-dependent) range of chemical potentials $\mu$ and temperatures $T$ (compare \Rcite{Buballa:2020nsi}). 
Then, an inhomogeneous phase is observed in the phase diagram using numerical minimization (for details regarding the algorithm we refer to Sec.~4.3 of \Rcite{Pannullo:2021edr}).
\begin{figure}[t]
	\centering
	\includegraphics[width=11.87cm]{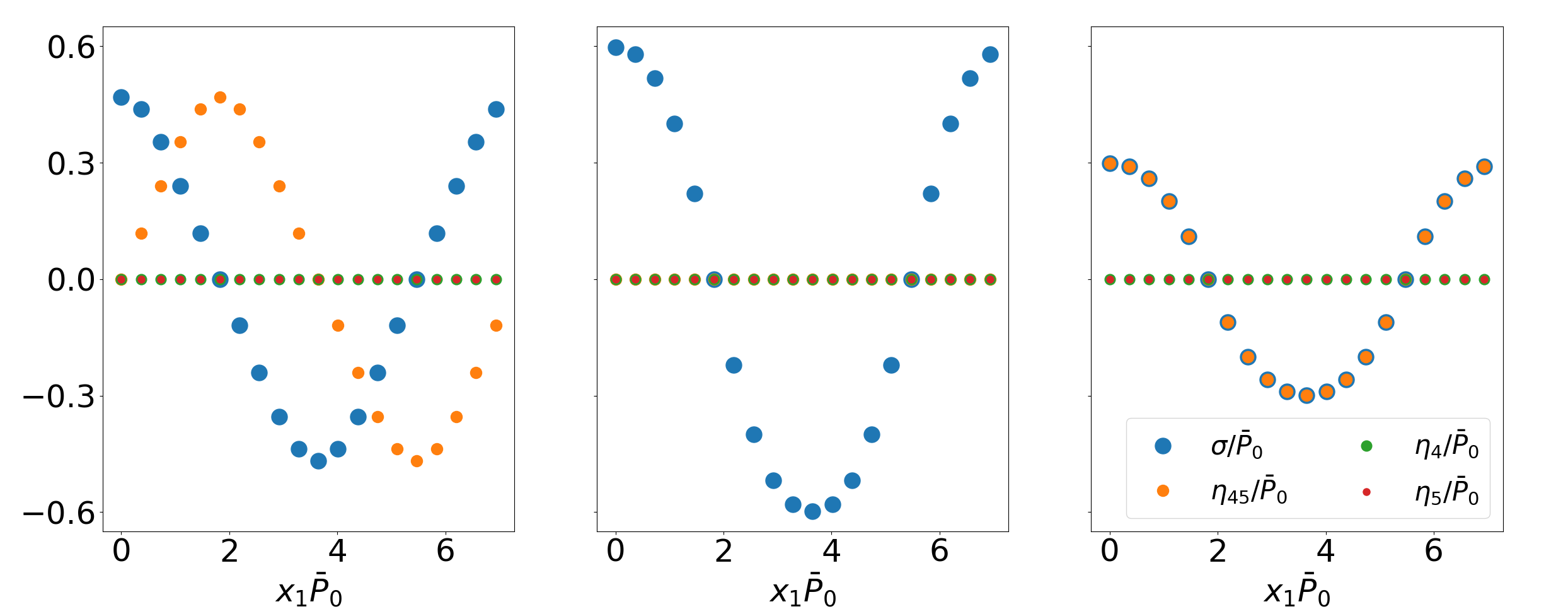}
	\caption{\label{fig:min}Inhomogeneous ground states using naive fermions at temperature $T/\bar P_0 = 0.137$, chemical potential $\mu/ \bar P_0 = 1.033$, finite lattice spacing $a\bar P_0 = 0.365$ and spatial volume $V\bar P_0^2 = (7.3)^2$. The three plotted minima are energetically degenerate up to the precision of our numerical minimization. The corresponding inhomogeneous phase is expected to vanish when taking $a\rightarrow 0$.}
\end{figure} 
We present an exemplary plot of inhomogeneous condensates obtained at finite $\mu/\bar P_0 = 1.033$ and $T/\bar P_0=0.137$ using naive fermions at finite lattice spacing $a\bar P_0 = 0.365$ in Fig.~\ref{fig:min}\footnote{Note that the naive discretization requires a weighting function with certain properties. This is, e.g., discussed in \cite{Cohen:1983nr, Lenz:2020bxk, Buballa:2020nsi}.}. 
Here, $\bar P_0=\sqrt{\bar{\Sigma}^2 + \bar{H}_4^2+ \bar{H}_5^2+ \bar{H}_{45}^2}\,\big|_{T=\mu=0}$ is used to set the scale, where $\bar \Sigma,\bar H_i$ are field configurations of $\sigma, \eta_i$ corresponding to the homogeneous, global minimum of the effective action. 
One obtains multiple degenerate inhomogeneous minima even when neglecting the ones which are related via global $O(3)$ rotations of the vector $\left(\sigma(\mathbf{x}), \eta_4(\mathbf{x}), \eta_5(\mathbf{x})\right)$. 
Note that the bosonic fields are allowed to be functions of the two spatial coordinates $\mathbf{x} = (x_1, x_2)$, but one-dimensional functions are observed as the resulting ground state after minimization of the effective action similar to our results in the GN model \cite{Pannullo:2021edr}.
We find degenerate minima, where both $\Sigma(\mathbf{x})$ and $H_{45}(\mathbf{x})$ oscillate. 
On the rightmost plot, they oscillate with the same phase, while on the leftmost plot they have a phase shift as observed for chiral density waves (see, e.g., \Rcite{Buballa:2014tba,Kutschera:1989yz, Schon:2000he}).
In the middle plot, only $\Sigma(\mathbf{x})$ oscillates while $H_4(\mathbf{x}) = H_5(\mathbf{x}) = H_{45}(\mathbf{x}) = 0$, but we note, again, that the oscillation can be shifted into $H_4(\mathbf{x})$ and/or $H_5(\mathbf{x})$ via the mentioned global $O(3)$ rotations. 
 
The external parameters $\mu$ and $T$ are in the order of the cutoff scale\footnote{Additionally, studying the model with the same finite cutoff for both the vacuum and at external parameters $\mu$ and $T$ violates renormalization group consistency, as discussed in \Rcite{Braun:2018svj}. This hints towards significant cutoff artifacts on the observed ground states.} $\tfrac{\pi}{a}$ and it is, thus, expected that the found ground state will change when varying $a$. 
Since the found inhomogeneities coincide with the region of negative $\Gamma^{(2)}_{\phi_i}(\mathbf{q}\neq 0)$ when evaluating the two-point functions \eqref{eq:2point_76} on the lattice (as in \Rcite{Buballa:2020nsi, Pannullo:2021edr}), the observed ground states are only cutoff effects and the inhomogeneous phase is expected to vanish in the continuum limit. 
We highlight that we do not observe inhomogeneous condensates on the lattice when $\Gamma^{(2)}_{\phi_i} \geq 0, \, \forall i$ at given $\mu$ and $T$, similar to \cite{Buballa:2020nsi, Pannullo:2021edr}.
\subsection{Nambu-Jona-Lasinio model in $2+1$ dimensions}
A FF model, which also features interactions in the isospin channel similar to the $3+1$-dimensional Nambu-Jona-Lasinio model, is given by 
	\begin{align}
\frac{S_{\text{eff}}[\sigma, \vec{\pi}_4, \vec{\pi}_5]}{\N}= \int \mathrm{d}^{3}x &\tfrac{\sigma^2 + \vec{\pi}_4^2 + \vec{\pi}_5^2}{2 \lambda} - \mathrm{Tr} \ln\left(\slashed{\partial}  + \gamma_0\mu + \sigma + \ii \gamma_{4} \vec{\tau}\, \vec{\pi}_4 + \ii \gamma_{5} \vec{\tau}\, \vec{\pi}_5\right)
\end{align}
with auxiliary fields $\sigma, \vec{\pi}_4, \vec{\pi}_5$.
Restricting to homogeneous condensates one can study only homogeneous minima with $\left(\sigma, \vec{\pi}_4, \vec{\pi}_5\right) = \left(\bar{\Sigma}, 0, 0\right)$ using chiral symmetry transformations. 
The second order correction due to the inhomogeneous perturbations is given by  
\begin{align}
&\frac{S_{\text{eff}}^{(2)}}{\N} = \tfrac{\beta}{4} \int \frac{\mathrm{d}^2q}{(2\uppi)^2} \sum_{\phi\in \{\sigma,  \vec{\pi}_4, \vec{\pi}_5\}} |\delta\tilde\phi(\mathbf{q})|^2 \, \Gamma_\phi^{(2)}(\mathbf{q}^2), \\
\Gamma^{(2)}_\sigma &=  \frac{1}{2\lambda}-\ell_1+ L_2(\mathbf{q}, \bar{\sigma}, \mu), \label{eq:GNJL} \quad \Gamma^{(2)}_{\vec{\pi}_5} = \Gamma^{(2)}_{\vec{\pi}_4} =  \frac{1}{2\lambda}-\ell_1-\frac{1}{2}\mathbf{q}^2\ell_2(\mathbf{q}^2, \bar{\sigma}, \mu), 
\end{align}
which again yields monotonically increasing two-point functions of $q$. The homogeneous expansion points $\left(\sigma, \vec{\pi}_4, \vec{\pi}_5\right) = \left(\bar{\Sigma}, 0, 0\right)$ are stable against inhomogeneous perturbations.
\subsection{The effect of chiral imbalance}
A chiral imbalance, described by a finite value of $\mu_{45}$, in general complicates the analysis. We study the effect on the $\mathrm{GN}_P$-model, given by the effective action 
\begin{align}
\frac{S_{\text{eff}}[\sigma, \eta_{45}]}{\N}= \int \mathrm{d}^{3}x &\tfrac{\sigma^2 + \eta_{45}^2}{2 \lambda} - \mathrm{Tr} \ln \left(\slashed{\partial}  + \gamma_0\mu + \gamma_0 \gamma_{45} \mu_{45} + \sigma + \gamma_{45} \eta_{45}  \right).
\end{align}
Minimizing the effective action for homogeneous condensates, i.e., finding $(\sigma, \eta_{45}) = \left(\bar{\Sigma}, \bar{H}_{45}\right)$, yields a non-vanishing $\bar{H}_{45}$ at finite $\mu_{45}$. This leads to off-diagonal terms in the second order contribution by the fermionic determinant, where $\Gamma_{\sigma \eta_{45}}^F \neq 0$.
This is resolved by a change of basis in field variables and chemical potentials using
\begin{equation}
	P_{L/R} = (\mathds{1} \pm \gamma_{45}), \quad
	\mu_{L/R} = (\mu \pm \mu_{45}), \quad \phi_{L/R} = (\sigma \pm \eta),
\end{equation}	
which yields
\begin{align}
\frac{S_{\text{eff}}[\phi_L, \phi_{R}]}{\N}= \int \mathrm{d}^{3}x &\frac{\phi_L^2 + \phi_R^2}{4 \lambda} - \mathrm{Tr} \ln \left(\slashed{\partial}  + \gamma_0\left(P_L\mu_L + P_R \mu_R\right)  + P_L \phi_L + P_R \phi_R \right).
\end{align}
The corresponding second order correction due to the inhomogeneous perturbations is given by 
\begin{align}
&\frac{S_{\text{eff}}^{(2)}}{\N} = \tfrac{\beta}{4} \int \frac{\mathrm{d}^2q}{(2\uppi)^2} \sum_{\phi\in \{\phi_L,  \phi_R\}} |\delta\tilde\phi(\mathbf{q})|^2\, \Gamma_\phi^{(2)}(\mathbf{q}^2), \\
\Gamma^{(2)}_{\phi_L} &=  \frac{1}{4\lambda}-\tfrac{1}{2}\ell_1+ \tfrac{1}{2}L_2(\mathbf{q}, \bar{\phi}_L, \mu_L),\quad \Gamma^{(2)}_{\phi_R} =  \frac{1}{4\lambda}-\tfrac{1}{2}\ell_1+ \tfrac{1}{2}L_2(\mathbf{q}, \bar{\phi}_{R}, \mu_R), 
\end{align}
where it becomes evident that this model decomposes into two independent GN models, each with one chemical potential $\mu_L/\mu_{R}$ and scalar field $\phi_{L}/\phi_{R}$ respectively. Consequently, the two-point functions do not signal the existence of an inhomogeneous phase in the $\mathrm{GN}_P$-model with chiral imbalance, as the homogeneous expansion points $\left(\sigma, \eta_{45}\right) = \left(\bar{\Sigma}, \bar{H}_{45}\right)$ are stable. 
\section{Summary}
We analyzed the stability of homogeneous chiral condensates against inhomogeneous perturbations in a variety of FF models, which cover different interaction channels and underlying (chiral) symmetry groups. In each of the studied models no indications for an inhomogeneous phase are observed. We found substantial arguments, that also Yukawa models, that correspond to the bosonized FF models via addition of kinetic terms and $2n$-self interactions for the auxiliary bosonic fields, do not develop an instability towards an inhomogeneous condensate.

\section*{Acknowledgments}
	We thank M.~Wagner for his encouragement to realize this project and furthermore for valuable comments, ideas and discussions.
	We further acknowledge useful discussions with M.~Buballa, A.~Koenigstein, L.~Kurth, J.~J.~Lenz, M.~Mandl, R.~D.~Pisarski, M.~J.~Steil and A.~Wipf.
	We thank A.~D'Ambrosio, A.~Koenigstein,  M.~Mandl and M.~Wagner for valuable comments on this manuscript.
	M.~Winstel thanks the organizers of "The 39th International Symposium on Lattice Field Theory" for the opportunity to give this talk.
	This work was supported by the Deutsche Forschungsgemeinschaft (DFG, German Research Foundation) – project number 315477589 – TRR 211.
	M.~Winstel acknowledges support by the GSI Forschungs- und Entwicklungsvereinbarungen (GSI F\&E).
	M.~Winstel acknowledges support of the Giersch Foundation.
	Calculations on the GOETHE-HLR and on the on the FUCHS-CSC high-performance computers of the Frankfurt University were conducted for this research.
	We would like to thank HPC-Hessen, funded by the State Ministry of Higher Education, Research and the Arts, for programming advice.

\bibliographystyle{JHEP}	
\bibliography{literature}

\providecommand{\href}[2]{#2}\begingroup\raggedright\begin{thebibliography}{10}

\bibitem{Gross:1974jv}
D.J.~Gross and A.~Neveu, \emph{{Dynamical Symmetry Breaking in Asymptotically
  Free Field Theories}},
  \href{https://doi.org/10.1103/PhysRevD.10.3235}{\emph{Phys. Rev. D}
  {\bfseries 10} (1974) 3235}.

\bibitem{Thies:2003kk}
M.~Thies and K.~Urlichs, \emph{{Revised phase diagram of the Gross-Neveu
  model}}, \href{https://doi.org/10.1103/PhysRevD.67.125015}{\emph{Phys. Rev.
  D} {\bfseries 67} (2003) 125015}
  [\href{https://arxiv.org/abs/hep-th/0302092}{{\ttfamily hep-th/0302092}}].

\bibitem{Thies:2006ti}
M.~Thies, \emph{{From relativistic quantum fields to condensed matter and back
  again: Updating the Gross-Neveu phase diagram}},
  \href{https://doi.org/10.1088/0305-4470/39/41/S04}{\emph{J. Phys. A}
  {\bfseries 39} (2006) 12707}
  [\href{https://arxiv.org/abs/hep-th/0601049}{{\ttfamily hep-th/0601049}}].

\bibitem{Schon:2000he}
V.~Schon and M.~Thies, \emph{{Emergence of Skyrme crystal in Gross-Neveu and 't
  Hooft models at finite density}},
  \href{https://doi.org/10.1103/PhysRevD.62.096002}{\emph{Phys. Rev. D}
  {\bfseries 62} (2000) 096002}
  [\href{https://arxiv.org/abs/hep-th/0003195}{{\ttfamily hep-th/0003195}}].

\bibitem{Thies:2020ofv}
M.~Thies, \emph{{Duality study of the chiral Heisenberg-Gross-Neveu model in
  1+1 dimensions}},
  \href{https://doi.org/10.1103/PhysRevD.102.096006}{\emph{Phys. Rev. D}
  {\bfseries 102} (2020) 096006}
  [\href{https://arxiv.org/abs/2008.13119}{{\ttfamily 2008.13119}}].

\bibitem{Ebert:2011rg}
D.~Ebert, N.V.~Gubina, K.G.~Klimenko, S.G.~Kurbanov and V.C.~Zhukovsky,
  \emph{{Chiral density waves in the NJL$_2$ model with quark number and
  isospin chemical potentials}},
  \href{https://doi.org/10.1103/PhysRevD.84.025004}{\emph{Phys. Rev. D}
  {\bfseries 84} (2011) 025004}
  [\href{https://arxiv.org/abs/1102.4079}{{\ttfamily 1102.4079}}].

\bibitem{Thies:2022kuv}
M.~Thies, \emph{{Tricritical curve of massive chiral Gross-Neveu model with
  isospin}}, \href{https://doi.org/10.1103/PhysRevD.106.056026}{\emph{Phys.
  Rev. D} {\bfseries 106} (2022) 056026}
  [\href{https://arxiv.org/abs/2207.14503}{{\ttfamily 2207.14503}}].

\bibitem{Khunjua:2017khh}
T.G.~Khunjua, K.G.~Klimenko, R.N.~Zhokhov and V.C.~Zhukovsky,
  \emph{{Inhomogeneous charged pion condensation in chiral asymmetric dense
  quark matter in the framework of NJL$_2$ model}},
  \href{https://doi.org/10.1103/PhysRevD.95.105010}{\emph{Phys. Rev. D}
  {\bfseries 95} (2017) 105010}
  [\href{https://arxiv.org/abs/1704.01477}{{\ttfamily 1704.01477}}].

\bibitem{Buballa:2014tba}
M.~Buballa and S.~Carignano, \emph{{Inhomogeneous chiral condensates}},
  \href{https://doi.org/10.1016/j.ppnp.2014.11.001}{\emph{Prog. Part. Nucl.
  Phys.} {\bfseries 81} (2015) 39}
  [\href{https://arxiv.org/abs/1406.1367}{{\ttfamily 1406.1367}}].

\bibitem{Lakaschus:2020caq}
P.~Lakaschus, M.~Buballa and D.H.~Rischke, \emph{{Competition of inhomogeneous
  chiral phases and two-flavor color superconductivity in the NJL model}},
  \href{https://doi.org/10.1103/PhysRevD.103.034030}{\emph{Phys. Rev. D}
  {\bfseries 103} (2021) 034030}
  [\href{https://arxiv.org/abs/2012.07520}{{\ttfamily 2012.07520}}].

\bibitem{Buballa:2020xaa}
M.~Buballa, S.~Carignano and L.~Kurth, \emph{{Inhomogeneous phases in the
  quark-meson model with explicit chiral-symmetry breaking}},
  \href{https://doi.org/10.1140/epjst/e2020-000101-x}{\emph{Eur. Phys. J. ST}
  {\bfseries 229} (2020) 3371}
  [\href{https://arxiv.org/abs/2006.02133}{{\ttfamily 2006.02133}}].

\bibitem{Carignano:2019ivp}
S.~Carignano and M.~Buballa, \emph{{Inhomogeneous chiral condensates in
  three-flavor quark matter}},
  \href{https://doi.org/10.1103/PhysRevD.101.014026}{\emph{Phys. Rev. D}
  {\bfseries 101} (2020) 014026}
  [\href{https://arxiv.org/abs/1910.03604}{{\ttfamily 1910.03604}}].

\bibitem{Buballa:2018hux}
M.~Buballa and S.~Carignano, \emph{{Inhomogeneous chiral phases away from the
  chiral limit}},
  \href{https://doi.org/10.1016/j.physletb.2019.02.045}{\emph{Phys. Lett. B}
  {\bfseries 791} (2019) 361}
  [\href{https://arxiv.org/abs/1809.10066}{{\ttfamily 1809.10066}}].

\bibitem{Nickel:2009wj}
D.~Nickel, \emph{{Inhomogeneous phases in the Nambu-Jona-Lasino and quark-meson
  model}}, \href{https://doi.org/10.1103/PhysRevD.80.074025}{\emph{Phys. Rev.
  D} {\bfseries 80} (2009) 074025}
  [\href{https://arxiv.org/abs/0906.5295}{{\ttfamily 0906.5295}}].

\bibitem{Dumm:2021vop}
D.G.~Dumm, J.P.~Carlomagno and N.N.~Scoccola, \emph{{Strong-interaction matter
  under extreme conditions from chiral quark models with nonlocal separable
  interactions}}, \href{https://doi.org/10.3390/sym13010121}{\emph{Symmetry}
  {\bfseries 13} (2021) 121}
  [\href{https://arxiv.org/abs/2101.09574}{{\ttfamily 2101.09574}}].

\bibitem{Andersen:2018osr}
J.O.~Andersen and P.~Kneschke, \emph{{Chiral density wave versus pion
  condensation at finite density and zero temperature}},
  \href{https://doi.org/10.1103/PhysRevD.97.076005}{\emph{Phys. Rev. D}
  {\bfseries 97} (2018) 076005}
  [\href{https://arxiv.org/abs/1802.01832}{{\ttfamily 1802.01832}}].

\bibitem{Lenz:2020bxk}
J.~Lenz, L.~Pannullo, M.~Wagner, B.~Wellegehausen and A.~Wipf,
  \emph{{Inhomogeneous phases in the Gross-Neveu model in 1+1 dimensions at
  finite number of flavors}},
  \href{https://doi.org/10.1103/PhysRevD.101.094512}{\emph{Phys. Rev. D}
  {\bfseries 101} (2020) 094512}
  [\href{https://arxiv.org/abs/2004.00295}{{\ttfamily 2004.00295}}].

\bibitem{Stoll:2021ori}
J.~Stoll, N.~Zorbach, A.~Koenigstein, M.J.~Steil and S.~Rechenberger,
  \emph{{Bosonic fluctuations in the $( 1 + 1 )$-dimensional
  Gross-Neveu(-Yukawa) model at varying $\mu$ and $T$ and finite $N$}},
  \href{https://arxiv.org/abs/2108.10616}{{\ttfamily 2108.10616}}.

\bibitem{Ciccone:2022zkg}
R.~Ciccone, L.~Di~Pietro and M.~Serone, \emph{{Inhomogeneous Phase of the
  Chiral Gross-Neveu Model}},
  \href{https://doi.org/10.1103/PhysRevLett.129.071603}{\emph{Phys. Rev. Lett.}
  {\bfseries 129} (2022) 071603}
  [\href{https://arxiv.org/abs/2203.07451}{{\ttfamily 2203.07451}}].

\bibitem{Lenz:2021kzo}
J.J.~Lenz, M.~Mandl and A.~Wipf, \emph{{Inhomogeneities in the two-flavor
  chiral Gross-Neveu model}},
  \href{https://doi.org/10.1103/PhysRevD.105.034512}{\emph{Phys. Rev. D}
  {\bfseries 105} (2022) 034512}
  [\href{https://arxiv.org/abs/2109.05525}{{\ttfamily 2109.05525}}].

\bibitem{Lenz:2021vdz}
J.J.~Lenz and M.~Mandl, \emph{{Remnants of large-$N_\mathrm{f}$ inhomogeneities
  in the 2-flavor chiral Gross-Neveu model}},
  \href{https://doi.org/10.22323/1.396.0415}{\emph{PoS} {\bfseries LATTICE2021}
  (2022) 415} [\href{https://arxiv.org/abs/2110.12757}{{\ttfamily
  2110.12757}}].

\bibitem{Nonaka:2021pwm}
C.~Nonaka and K.~Horie, \emph{{Inhomogeneous phases in the chiral Gross-Neveu
  model on the lattice}}, \href{https://doi.org/10.22323/1.396.0150}{\emph{PoS}
  {\bfseries LATTICE2021} (2022) 150}
  [\href{https://arxiv.org/abs/2112.02261}{{\ttfamily 2112.02261}}].

\bibitem{Winstel:2019zfn}
M.~Winstel, J.~Stoll and M.~Wagner, \emph{{Lattice investigation of an
  inhomogeneous phase of the 2 + 1-dimensional Gross-Neveu model in the limit
  of infinitely many flavors}},
  \href{https://doi.org/10.1088/1742-6596/1667/1/012044}{\emph{J. Phys. Conf.
  Ser.} {\bfseries 1667} (2020) 012044}
  [\href{https://arxiv.org/abs/1909.00064}{{\ttfamily 1909.00064}}].

\bibitem{Buballa:2020nsi}
M.~Buballa, L.~Kurth, M.~Wagner and M.~Winstel, \emph{{Regulator dependence of
  inhomogeneous phases in the ( 2+1 )-dimensional Gross-Neveu model}},
  \href{https://doi.org/10.1103/PhysRevD.103.034503}{\emph{Phys. Rev. D}
  {\bfseries 103} (2021) 034503}
  [\href{https://arxiv.org/abs/2012.09588}{{\ttfamily 2012.09588}}].

\bibitem{Narayanan:2020uqt}
R.~Narayanan, \emph{{Phase diagram of the large $N$ Gross-Neveu model in a
  finite periodic box}},
  \href{https://doi.org/10.1103/PhysRevD.101.096001}{\emph{Phys. Rev. D}
  {\bfseries 101} (2020) 096001}
  [\href{https://arxiv.org/abs/2001.09200}{{\ttfamily 2001.09200}}].

\bibitem{Winstel:2021yok}
M.~Winstel, L.~Pannullo and M.~Wagner, \emph{{Phase diagram of the
  2+1-dimensional Gross-Neveu model with chiral imbalance}},
  \href{https://doi.org/10.22323/1.396.0381}{\emph{PoS} {\bfseries LATTICE2021}
  (2022) 381} [\href{https://arxiv.org/abs/2109.04277}{{\ttfamily
  2109.04277}}].

\bibitem{Pannullo:2021edr}
L.~Pannullo, M.~Wagner and M.~Winstel, \emph{{Inhomogeneous Phases in the
  Chirally Imbalanced 2 + 1-Dimensional Gross-Neveu Model and Their Absence in
  the Continuum Limit}},
  \href{https://doi.org/10.3390/sym14020265}{\emph{Symmetry} {\bfseries 14}
  (2022) 265} [\href{https://arxiv.org/abs/2112.11183}{{\ttfamily
  2112.11183}}].

\bibitem{Gies:2009da}
H.~Gies, L.~Janssen, S.~Rechenberger and M.M.~Scherer, \emph{{Phase transition
  and critical behavior of d=3 chiral fermion models with left/right
  asymmetry}}, \href{https://doi.org/10.1103/PhysRevD.81.025009}{\emph{Phys.
  Rev. D} {\bfseries 81} (2010) 025009}
  [\href{https://arxiv.org/abs/0910.0764}{{\ttfamily 0910.0764}}].

\bibitem{Pisarski:1984dj}
R.D.~Pisarski, \emph{{Chiral Symmetry Breaking in Three-Dimensional
  Electrodynamics}},
  \href{https://doi.org/10.1103/PhysRevD.29.2423}{\emph{Phys. Rev. D}
  {\bfseries 29} (1984) 2423}.

\bibitem{Koenigstein:2021llr}
A.~Koenigstein, L.~Pannullo, S.~Rechenberger, M.J.~Steil and M.~Winstel,
  \emph{{Detecting inhomogeneous chiral condensation from the bosonic two-point
  function in the (1 + 1)-dimensional Gross\textendash{}Neveu model in the
  mean-field approximation*}},
  \href{https://doi.org/10.1088/1751-8121/ac820a}{\emph{J. Phys. A} {\bfseries
  55} (2022) 375402} [\href{https://arxiv.org/abs/2112.07024}{{\ttfamily
  2112.07024}}].

\bibitem{Nakano:2004cd}
E.~Nakano and T.~Tatsumi, \emph{{Chiral symmetry and density wave in quark
  matter}}, \href{https://doi.org/10.1103/PhysRevD.71.114006}{\emph{Phys. Rev.
  D} {\bfseries 71} (2005) 114006}
  [\href{https://arxiv.org/abs/hep-ph/0411350}{{\ttfamily hep-ph/0411350}}].

\bibitem{Klimenko:1987gi}
K.G.~Klimenko, \emph{{Phase Structure of Generalized {Gross-Neveu} Models}},
  \href{https://doi.org/10.1007/BF01578141}{\emph{Z. Phys. C} {\bfseries 37}
  (1988) 457}.

\bibitem{Pannullo:Lattice2022}
L.~Pannullo, M.~Wagner and M.~Winstel, \emph{{Inhomogeneous phases in the
  3+1-dimensional mean-field Nambu-Jona-Lasinio model on the lattice}},
  {\emph{in preparation} (2022) }.

\bibitem{Cohen:1983nr}
Y.~Cohen, S.~Elitzur and E.~Rabinovici, \emph{{A MONTE CARLO STUDY OF THE
  GROSS-NEVEU MODEL}},
  \href{https://doi.org/10.1016/0550-3213(83)90136-0}{\emph{Nucl. Phys. B}
  {\bfseries 220} (1983) 102}.

\bibitem{Kutschera:1989yz}
M.~Kutschera, W.~Broniowski and A.~Kotlorz, \emph{{Quark Matter With Neutral
  Pion Condensate}},
  \href{https://doi.org/10.1016/0370-2693(90)91421-7}{\emph{Phys. Lett. B}
  {\bfseries 237} (1990) 159}.

\bibitem{Braun:2018svj}
J.~Braun, M.~Leonhardt and J.M.~Pawlowski, \emph{{Renormalization group
  consistency and low-energy effective theories}},
  \href{https://doi.org/10.21468/SciPostPhys.6.5.056}{\emph{SciPost Phys.}
  {\bfseries 6} (2019) 056} [\href{https://arxiv.org/abs/1806.04432}{{\ttfamily
  1806.04432}}].

\end{thebibliography}\endgroup
\end{document}